\documentclass[useAMS,usenatbib]{mn2e}
\usepackage{graphicx,epsfig,psfig}
\usepackage{url}

\title[Metallicity Distributions in Cosmological Disc Simulations]{The Distribution Of Metals in Cosmological Hydrodynamical Simulations of Dwarf Disk Galaxies}
\author[K. Pilkington et~al.]
  {K.~Pilkington$^{1,2,3}$,
   B.K.~Gibson$^{1,2,3}$,
   C.B.~Brook$^{1,4}$,\newauthor
   F.~Calura$^{1,5}$,
   G.S.~Stinson$^{1,6}$, 
   R.J.~Thacker$^2$, 
   L.~Michel-Dansac$^7$,\newauthor
   J.~Bailin$^8$,
   H.M.P.~Couchman$^9$,
   J.~Wadsley$^9$,
   T.R.~Quinn$^{10}$, and
   A.~Maccio$^{6}$\\
$^{1}$Jeremiah Horrocks Institute, University of Central Lancashire,
Preston, PR1~2HE, UK\\
$^{2}$Department of Astronomy \& Physics, Saint Mary's University, Halifax,
Nova Scotia, B3H~3C3, Canada\\
$^{3}$Monash Centre for Astrophysics, School of Mathematical Sciences,
Monash University, Clayton, VIC, 3800, Australia\\
$^{4}$Departamento de F\'isica Te\'orica, Universidad Aut\'onoma de Madrid,
E-28049 Cantoblanco, Madrid, Spain\\
$^{5}$Istituto Nazional di Astrofisica, Osservatorio Astronomico di Bologna, 
Via Ranzani 1, I-40127, Bologna, Italy\\
$^{6}$Max-Planck-Institut f\"ur Astronomie, K\"onigstuhl 17, 69117
Heidelberg, Germany\\
$^{7}$Centre de Recherche Astrophysique de Lyon, Universit\'e de Lyon,
Obs. de Lyon, CNRS, Saint-Genis Laval, 69230, France\\
$^{8}$Astronomy Department, University of Michigan, 500 Church St.,
Ann Arbor, MI, 48109-1042, USA\\
$^{9}$Department of Physics \& Astronomy, McMaster University, Hamilton,
Ontario, L8S~4M1, Canada\\
$^{10}$Astronomy Department, University of Washington, Box 351580, Seattle,
WA, 98195, USA}

\begin{document}
\date{Submitted}
\pagerange{\pageref{firstpage}--\pageref{lastpage}} 
\pubyear{2012}

\maketitle
\label{firstpage}

\begin{abstract}
We examine the chemical properties of five cosmological hydrodynamical 
simulations of an M33-like disc galaxy which have been shown previously 
to be consistent with the morphological characteristics and bulk scaling 
relations expected of late-type spirals.  These simulations 
are part of the Making Galaxies In a Cosmological Context (MaGICC)
Project, in which stellar feedback is tuned to match the 
stellar mass -- halo mass relationship. Each realisation employed 
identical initial conditions and assembly histories, but differed from 
one another in their underlying baryonic physics prescriptions, 
including (a) the efficiency with which each supernova energy couples to 
the surrounding interstellar medium, (b) the impact of feedback 
associated with massive star radiation pressure, (c) the role of the 
minimum shut-off time for radiative cooling of Type~II supernovae 
remnants, (d) the treatment of metal diffusion, and (e) varying the 
initial mass function.  Our analysis focusses on the resulting stellar 
metallicity distribution functions (MDFs) in each simulated (analogous) 
`solar neighbourhood' (2$-$3 disc scalelengths from the galactic centre) 
and central `bulge' region. We compare and contrast the simulated MDFs' 
skewness, kurtosis, and dispersion (inter-quartile, inter-decile, 
inter-centile, and inter-tenth-percentile regions) with that of the 
empirical solar neighbourhood MDF and local group dwarf galxies.  
We find that the MDFs of the 
simulated discs are more negatively skewed, with higher kurtosis, than 
those observed locally in the Milky Way and local group dwarfs.  
We can trace this difference 
to the simulations' very tight and correlated age-metallicity relations 
(compared with that of the Milky Way's solar neighbourhood), suggesting 
that these relations within `dwarf' discs might be steeper than in 
L$_\star$ discs (consistent with the simulations' star formation 
histories and extant empirical data) and/or the degree of stellar 
orbital re-distribution and migration inferred locally has not been 
captured in their entirety, at the resolution of our simulations. The 
important role of metal diffusion in ameliorating the over-production of 
extremely metal-poor stars is highlighted.
\end{abstract}


\section{Introduction}
\label{intro}

The relative number of stars of a given metallicity in a given 
environment, whether it be the local stellar disc, central 
spheroid/bulge, and or baryonic halo -- the so-called metallicity 
distribution function (MDF) -- has embedded within it, the time 
evolution of a system's star formation, assembly/infall, and outflow 
history, all convolved with the initial mass function (IMF) 
\citep{Tinsley1980}. Seminal reviews of the diagnostic power of the MDF 
include those of \citet{Haywood2001} and \citet{Caimmi2008}.

Well in advance of our now empirical appreciation of (a) the 
hierarchical assembly of galaxies from sub-galactic units, (b) the 
ongoing infall of fresh material from halos to discs (e.g. High-Velocity 
Clouds: \citet{Gibson2001}), and (c) the ongoing outflow of enriched 
material from discs via stellar- and supernovae-driven winds/fountains 
\citep[e.g.][]{McClure2006}, it was recognised that the local MDF 
provided crucial evidence that the Milky Way (and presumably galaxies as 
a whole) did not behave as a `closed-box', in an evolutionary sense 
\citep{Pagel1975}.

This latter recognition was perhaps best manifest in what became known 
as the `G-dwarf Problem' \citep{Hartwick1976}; specifically, a simple 
model in which gas was not allowed to infall or outflow from the system 
would necessarily lead to a significant population of long-lived, low 
metallicity, stars in the solar neighbourhood, with $\sim$20\% of the 
stars locally predicted to possess metallicities below 
[Fe/H]$\approx$$-$1 \citep{Tinsley1980}.  In nature, such a population 
is not observed, with the empirical fraction of local low-metallicity 
stars being $\sim$2 orders of magnitude smaller than the aforementioned 
closed-box predictions \citep[e.g.][]{Kotoneva2002, Cas11}.

Since this recognition of its fundamental importance, the MDF has acted 
as one of the primary constraints / boundary conditions against which 
all analytical \citep[e.g.][]{Sch09, Kirby11}, semi-numerical 
\citep[e.g.][]{Chiap01, Fenner2003}, and chemo-dynamical 
\citep[e.g.][]{Rok08b, San09, Tissera2011, Calura2012} models are 
compared.

From a chemo-dynamical perspective, recent work has focused on the 
sensitivity of global metal re-distribution to different physical 
prescriptions, within the context of the OWLS project \citet{Wier11}; at 
higher redshift, a similar, equally comprehensive, study was undertaken 
by \citet{Som08}.  In both cases, the emphasis was placed on the 
whereabouts of the `missing metals' -- i.e., metals thought to reside in 
the Warm-Hot Intergalactic Medium (WHIM) and/or halos of massive 
galaxies, but have thus far proven challenging to detect 
directly.\footnote{\citet{Tumlinson2011} is an excellent example of 
recent efforts, though, to characterise the properties of these 
difficult-to-observe baryon reservoirs.}  While not fully cosmological, 
the reader is also referred to the excellent chemo-dynamical work of 
\citet{Kob11}, for a complementary analysis of a simulated Milky 
Way-like system.

Each of the above chemo-dynamical studies examine cursory aspects of the 
MDF `constraint', but the focus for each was never meant to be a 
comprehensive analysis of the dispersion and higher-order moments of the 
shape characteristics,\footnote{cf. \citet{Kirby11}, though, for a study 
of the higher-order moments of the MDFs of Local Group dwarf spheroidals 
which is similar in spirit to our work here on disc galaxies.} nor their 
link to the associated age-metallicity relations, star formation 
histories, and putative G-dwarf problem; such higher-order moments 
include the MDF skewness, kurtosis, and inter-quartile, inter-decile, 
inter-centile, and inter-tenth-percentile regions.

The skewness of an MDF can be a reflection of both the classical G-dwarf 
problem and the slope of the age-metallicity relation (AMR); kurtosis is 
often thought of as being a measure of the `peakedness' of the MDF 
(e.g., by how much the peak is `flatter' or `peakier' than a Gaussian), 
while in practice it is often more sensitive to the presence of `heavy' 
tails, rather than the shape of the peak; the inter-quartile, -decile, 
etc., regions probe both the effects of star formation histories and 
AMRs and, in the case of the inter-centile and (especially) the 
inter-tenth-percentile regions, the impact of metal diffusion on the 
extreme metal-poor tail of the distribution.  In the context of 
cosmological chemo-dynamical disc simulations, to our knowledge, our's 
is the first quantitative discussion of these higher-order moments of 
the MDF.

Further, from an observational perspective, the recent re-calibrations 
of the original Geneva-Copenhagen Survey (GCS: \citet{Nod04}) by 
\citet{Holm09} and \citet{Cas11} has made for a timely investigation of 
the predicted characteristics of the MDFs of simulated disc galaxies.  
Parallel developments slightly further afield\footnote{Spatially 
speaking, in relation to that of the solar neighbourhood region probed 
by the GCS.} include targeted MDF studies of the thin$-$thick disc 
transition region and the thick disk proper \citep{Schlesinger2012}, the 
stellar halo \citep{Sch09}, and the Galactic bulge 
\citep[][]{Ben11,Hill11}.

This paper aims to fill a gap in the literature, by making use of a new 
suite of fully cosmological chemo-dynamical simulations whose properties 
have been shown to be in remarkable agreement with the basic scaling 
laws to which late-type disc galaxies adhere in nature \citep{Brook11c, 
Maccio12}. The simulations themselves are outlined briefly in 
\S\ref{secgal}, alongside a description of the adopted analogous `solar 
neighbourhood' regions. The associated age-metallicity relations (AMRs) 
are presented in \S\ref{secagemet}; the need for this will become 
apparent when analysing the higher-order moments of the MDFs within 
these regions and, in particular, their metal-poor tails 
(\S\ref{secmdf}).  Our results will then be summarised in 
\S\ref{summary}.

\section{Simulations}
\label{secgal}

In what follows, we analyse five cosmological zoom variants of the 
`scaled-down' M33-like disc galaxy simulation (\tt g15784\rm) described 
by \citet{Brook11b}.  The initial conditions are identical for each 
realisation, and taken from the eponymous \tt g15784 \rm of 
\citet{Stinson10} after re-scaling \citep[e.g.][]{Kannan2012} the mass 
(length) scales by a factor of eight (two).
Differences in the underlying power spectrum that result from this 
re-scaling are minor 
\citep[e.g.][]{Springel2008,Maccio12,Kannan2012,VeraCiro2012},  
and do not affect our results. 
The virial mass of the 
scaled \tt g15784 \rm is 2$\times$10$^{11}$~M$_\odot$, with 
$\sim$10$^{7}$ particles within the virial radius at $z$=0, with a mean 
stellar particle mass of $\sim$6400~M$_\odot$.  A gravitational 
softening of $\epsilon$=155~pc was used; to ensure that gas resolves the 
Jeans mass, rather than undergoing artifical fragmentation, pressure is 
added to the gas, after \citet{Robertson2008}.  Further, a maximum 
density limit is imposed by setting the minimum SPH smoothing length to 
be 1/4 that of the softening length.\footnote{In comparison, the 
original \citet{Stinson10} simulations used a minimum SPH smoothing 
length of $\epsilon$/100, resulting a dramatic increase in computational 
time, but with only minimal impact on the simulation itself.}

Each of the five simulations was evolved using the gravitational N-body 
+ smoothed particle hydrodynamics (SPH) code \textsc{Gasoline} 
\citep{Wad04}.  Metal-dependent cooling of the gas, under the assumption 
of ionisation equilibrum, is applied, after \citet{SWS10}, coupled to a 
uniform, evolving, \citet{Haardt96} ionising ultraviolet background. Our 
reference/fiducial simulation (\tt 11mKroupa\rm) was introduced by 
\citet{Brook11b}, in the context of its outflow and angular momentum 
characteristics. The structural and kinematic properties 
(e.g., rotation curves, bulge-to-disc decomposition, ratio of 
rotational-to-anisotropic support, etc.) of 
the simulations presented here are indistinguishable from those
presented in \citet{Brook11b}, to which the reader is referred for
supplementary details.

When gas reaches a sufficiently cool temperature -- 
$T$$<$10,000$-$15,000~K -- and resides within a sufficiently dense 
environment -- $n_{\rm th}$$>$9.3~cm$^{-3}$ --\footnote{The star 
formation density threshold $n_{\rm th}$ corresponds to the maximum 
density gas can reach using gravity -- i.e., $n_{\rm th}$=32$M_{\rm 
gas}$/$\epsilon^3$.} it becomes eligible to form stars according to 
$\frac{dM_{\star}}{dt}$$=$c$_{\star}$$\frac{M_{gas}}{t_{dyn}}$, where 
c$_{\star}$ is the star formation efficiency,\footnote{The star 
formation efficiency $c_\star$ was taken to be 10\% for all the runs, 
except for \tt 11mChab\rm, for a value of 7.5\% was adopted.} $\Delta t$ 
is the timestep between star formation events (0.8~Myrs, here), 
$M_{gas}$ is the SPH particle mass, $t_{dyn}$ is the SPH particle's 
dynamical time, and $\Delta M_\star$ is the mass of the star particle 
formed.

We have extended the chemical `network' of \textsc{Gasoline} from oxygen 
and iron, to now also track the evolution of carbon, nitrogen, neon, 
magnesium, and silicon.  After \citet{Rait96}, power law fits to the 
\citet{WooWe95} Z=0.02 SNeII yields were generated for the dominant 
isotopes for each of these seven elements; a further extension was 
implemented, in order to include the \citet{Hoek97} 
metallicity-dependent carbon, nitrogen, and oxygen yields from 
asymptotic giant branch (AGB) stars.  By expanding upon the chemical 
species being tracked, the earlier concern regarding the underprediction 
of the global metallicity by a factor of $\sim$2 (and the consequent 
underestimate to the SPH cooling and star formation rates) is naturally 
alleviated \citep{Pilkington2012}. We note in passing that all 
abundances (and ratios) presented here are relative to the solar scale 
defined by \citet{Asp09}.

Feedback from supernovae (SNe) follows the blastwave formalism of 
\citet{Stinson06}, with 100\% of the energy (10$^{51}$~erg/SN) thermally 
coupled to the surrounding ISM.  Cooling is disabled for particles 
within the blast region (corresponding to the radius of the remnant when 
the interior pressure has been reduced to that of the pressure of the 
ambient ISM) for a timescale corresponding to that required to cool the 
hot interior gas to $T$$\sim$10$^4$~K.\footnote{To use the terminology 
of \citet{Gibson1994}, the relevant radius and timescale correspond to 
R$_{\rm merge}$ and t$_{\rm cool}$, respectively.}  Bearing in mind the 
0.8~Myr timesteps of our runs, we impose a minimum cooling `shut-off 
time' which matches this value.\footnote{Save, for the one run for which 
this restriction was relaxed (\tt 11mNoMinShut\rm).}

We employ the ``MaGICC'' (Making Galaxies In a Cosmological Context) 
feedback model described by \citet{Brook11b} and \citet{Stinson12}, 
taking into account the effect of energy
feedback from massive stars into 
the ISM\footnote{Except for the one run included here 
without radiation energy (\tt 11mNoRad\rm).} (cf. \citet{Hopkins2011}).  
While a typical massive star might emit $\sim$10$^{53}$~erg of radiation 
energy during its pre-SN lifetime, these photons do not couple 
efficiently to the surrounding ISM; as such, we only inject 10\% of this 
energy in the form of thermal energy into the surrounding gas, and 
cooling is not disabled for this form of energy input. Of this injected 
energy typically 90-100\% is radiated away within a single dynamical time.

The default initial mass function (IMF) is that of \citet{Kroupa1993}; 
the \tt 11mChab \rm run incorporates the more contemporary (and 
currently favoured) \citet{Chabrier2003} functional form; per stellar 
generation, the latter possesses a factor of $\sim$4$\times$ the number 
of SNeII as that of the former. Finally, the treatment of metal 
diffusion within \textsc{Gasoline} is detailed by \citet{SWS10}; a 
diffusion coefficient $C$=0.05 has been adopted for our runs, except for 
one simulation for which diffusion was prohibited (\tt 
11mNoDiff\rm).\footnote{Our `no diffusion' run possesses MDF and 
chemical `characteristics' similar to those of \tt DG1 \rm 
\citep{Gov10}, the latter for which a brief chemical analysis was shown 
in \citet{Pilk10}. This similarity can be traced to the less efficicient 
metal diffusion adopted for the \tt DG1 \rm runs (i.e., $C$=0.01 vs the 
$C$=0.05 now employed for our \textsc{Gasoline} runs, after 
\citet{SWS10}).}
The primary numerical characteristics of the five simulations employed
here are listed in Table~\ref{tab}.

\begin{table*}
\centering
\begin{tabular}{|l|l|l|r|r|r|r|c|c|c|}
\hline 
Galaxy             & IMF      & c$_\star $ & $\epsilon$SN & SR   & T$_{max}$ & Stellar Mass       & Scale Length & Vertical Gradient & Radial Gradient\\ \hline\hline
{\tt 11mKroupa}    & Kroupa   & 0.1        & 100\%        & 10\% & 15000     &  7.1$\times$10$^9$ & 2.34         & $-$0.064          & $-$0.012       \\ \hline
{\tt 11mChab}      & Chabrier & 0.075      & 100\%        & 10\% & 10000     &  1.3$\times$10$^9$ & 2.78         & $-$0.017          & $-$0.026       \\ \hline
{\tt 11mNoRad}     & Kroupa   & 0.1        & 100\%        &  0\% & 15000     &  9.1$\times$10$^9$ & 1.58         & $-$0.027          & $-$0.045       \\ \hline
{\tt 11mNoMinShut} & Kroupa   & 0.1        & 100\%        & 10\% & 15000     & 14.0$\times$10$^9$ & 1.71         & $-$0.008          & $-$0.020       \\ \hline
{\tt 11mNoDiff}    & Kroupa   & 0.1        & 100\%        & 10\% & 10000     &  2.1$\times$10$^9$ & 1.43         & $-$0.013          & $-$0.028       \\ \hline
\end{tabular}
\caption{Primary parameters employed for the five simulations analysed 
in this work.  Column (1): simulation/galaxy name; Column (2): adopted 
IMF (Kroupa$\equiv$\citet{Kroupa1993}; Chabrier$\equiv$\citet{Chabrier2003}; 
Column (3): star formation efficiency; Column (4): thermalised SN 
energy fraction coupled to the ISM; Column (5): thermalised massive star 
radiation energy fraction coupled to the ISM; Column (6): maximum 
allowable gas temperature for star formation; Column (7): present-day 
stellar mass (in solar masses) within the virial radius; Column (8): 
stellar disc exponential scalelength (in kpc); Column (9): vertical 
[Fe/H] gradient (in dex/kpc); Column (10): radial [Fe/H] gradient (in 
dex/kpc).}
\label{tab}
\end{table*}

For our MDF and AMR analyses, for each simulation we identify an 
analogous region to that of the Milky Way's `solar neighbourhood', 
defined to be a radial range from 3.0 to 3.5 disc scalelengths (see 
Table~\ref{tab}) and to lie within 500~pc of the galactic mid-plane. The 
fraction of accreted stars in these high-feedback runs is negligble; as 
such their contamination in the `solar neighbourhood' is equally 
negligible.  Consequently, there was no need to undertake the sort of 
kinematic decomposition of the orbital circularity $\epsilon_J\equiv 
J_z/J_{circ}(E)$ distribution\footnote{Where $J_z$ is the $z$-component 
of the specific angular momentum and $J_{circ}(E)$ is the angular 
momentum of a circular orbit at a given specific binding energy.} that 
was needed to isolate disc/in-situ stars from spheroid/accreted stars in 
our parallel analysis of the MDFs of the more massive (and 
accretion-contaminated) \citet{Stinson10} simulations 
\citep{Calura2012}.\footnote{Note, this was confirmed by undertaking a 
kinematic decomposition of \tt 11mKroupa \rm using the modified technique 
introduced by \citet{Abad03b}, and employed by \citet{Calura2012}; 
specifically, none of our conclusions were contingent upon the need for 
a kinematic decomposition.  More quantitatively, only $\sim$3\% of
the stars in our simulated `solar neighbourhoods' would be
kinematically classified as `bulge/spheroid' stars, impacting
on the various MDF metrics to be discussed later at the $<$3\%
level (smaller than the uncertainty associated with the 
treatment of extreme ($>$5$\sigma$) outliers - see \S4). 
In light of this negligible impact, we have avoided imposing
any personal preferred kinematic decomposition scheme into the
analysis.\rm}

\begin{figure}
\centerline{ 
\psfig{file=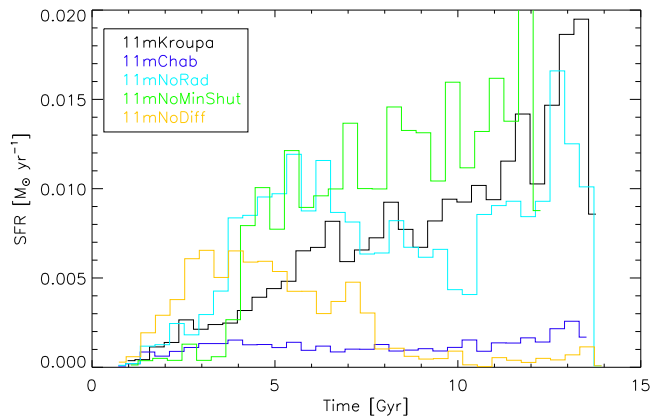,width=9.0cm} 
} 
\caption{Star formation histories of the solar neighbourhoods associated
with the five simulations; colour-coding is as noted in the
inset to the panel.}
\label{sfr}
\end{figure}

We first show the inferred star formation histories (SFHs) of the solar 
neighbourhoods associated with each of the five simulations 
(Fig~\ref{sfr}).  Several important points should be made, before 
analysing the AMRs and MDFs.  Qualitatively speaking, the SFHs of these 
regions within \tt 11mKroupa\rm, \tt 11mNoMinShut\rm, and \tt 11mNoRad 
\rm are similar to those seen in gas-rich dwarfs like NGC~6822, 
Sextans~A, WLM, and to some extent, the LMC \citep{Dolphin2005}.  In 
that sense, they are (not surprisingly) different from the typical 
exponentially-decaying SFH (timescales of $\sim$5$-$7~Gyrs) inferred for 
the Milky Way's solar neighbourhood \citep[e.g.][]{Renda2005}, and so we 
should not expect {\it identical \rm} trends in the ancillary AMRs and 
MDFs, as those observed locally.  Indeed, we will show this to be case 
momentarily, but our interest here is more in identifying trends, rather 
than {\it exact \rm} star-by-star comparisons.

The one simulation which shows an exponentially-declining 
SFH at later times is that of \tt 11mNoDiff\rm; the lack of 
diffusion here acts to minimise the `spread' of metals to a degree that 
star formation is restricted (preferentially) to much less enriched SPH 
particles (in part, because the cooling then becomes less efficient for 
a greater number of SPH particles, which has a greater impact at later 
times where there are fewer efficiently cooling metal-enriched SPH 
particles out of which to potentially form stars. We will return to the 
special case of the `no diffusion' model shortly.

The SFH of \tt 11mChab \rm also shows a distinct behaviour relative to 
the \tt 11mKroupa \rm fiducial.  Specifically, it is significantly 
lower, and relatively constant, at all times; in spirit, this is similar 
to the inferred SFH of the LMC \citep[e.g.][]{Holtzman1999}.  This is 
reflected in the stellar mass at $z$=0 being significantly lower than 
\tt 11mKroupa\rm, which in turn aids considerably in bringing its 
properties into close agreement with essentially all traditional scaling 
relations \cite{Brook11c}.  This behaviour is driven by (a) the factor 
of four increase in the SNe per stellar generation (via the more massive 
star-biased IMF), and (b) the reduced maximum temperature for star 
formation (as noted earlier).

The subtle effect of allowing the minimum shut-off time for radiative 
cooling of SN remnants to become prohibitively small in high-density 
regions (in practice what this means is that the shut-off time becomes 
smaller than the timestep of 0.8~Myrs) can be seen in the \tt 
11mNoMinShut \rm curve of Fig~\ref{sfr}.  Specifically, SPH particles 
affected by this effectively cool `instantly' within the same timestep, 
without any delay.  Hence, the particles in question become `available' 
for star formation much sooner than they might otherwise; this has the 
effect of `boosting' the star formation relative to that of the fiducial 
\tt 11mKroupa\rm.

\section{Age-Metallicity Relations}
\label{secagemet}

\begin{figure}
\centerline{ 
\psfig{file=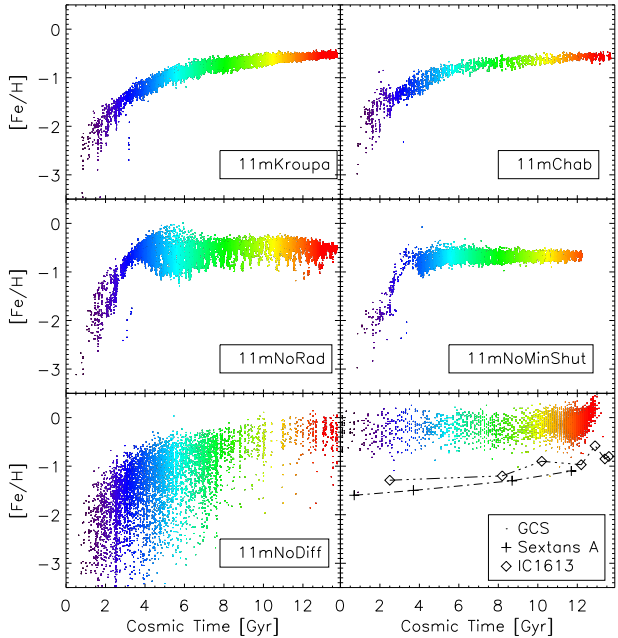,width=9.0cm} 
}
\caption{Age-metallicity relations (where metallicity$\equiv$[Fe/H]) in 
the analogous solar neighbourhoods of the five simulations employed 
here, in addition to the reference relationship found in the solar 
neighbourhood of the Milky Way and the dwarf irregulars Sextans~A 
\citep{Dolphin03} and IC~1613 \citep{Skillman2003}. Colour-coding in 
each panel is by stellar age, ranging from black/blue (oldest) to red 
(youngest).}
\label{agefe}
\end{figure}

As noted earlier, the MDF bears the imprint of a region's star formation 
history (SFH), convolved with its age-metallicity relation (AMR). Having 
introduced the `solar neighbourhood' SFHs in \S\ref{secgal}, we now 
present their associated AMRs in Fig~\ref{agefe}.  The time evolution of 
the [Fe/H] abundances are shown for each of the five simulations listed 
in Table~\ref{tab}.  Colour-coding within each panel corresponds to 
stellar age, ranging from old (black/blue) to young (red).

To provide a representative empirical dataset against which to compare, 
we make use of the recent re-calibration of the Geneva-Copenhagen Survey 
(GCS) presented bt \citet{Holm09}. The base GCS provides invaluable 
spectral parameters for $\sim$17,000 F- and G- stars in the solar 
neighbourhood.  Following \citet{Holm07}, we define a `cleaned' 
sub-sample by eliminating (i) binary stars, (ii) stars for which the 
uncertainty in age is $>$25\%, (iii) stars for which the uncertainty in 
trigonometric parallax is $>$13\%, and (iv) stars for which a `null' 
entry was provided for any of the parallax, age, metallicity, or their 
associated uncertainties.  The AMR for this `cleaned' sub-sampled of 
$\sim$4,000 stars is shown in the lower-right panel of Fig~\ref{agefe}.  
A fifth criterion is applied for the determination of the higher-order 
moments of the MDF shape; specifically, following \citet{Holm07} and 
constructing an unbiased volume-limited sub-sample from the stars lying 
within 40~pc of the Sun.  Doing so yields a smaller sample of only 
$\sim$500 stars.  While this does not impact on the shape 
characteristics of \S\ref{secmdf} or the behaviour of the AMR, for 
clarity, we show the AMR inferred from the aforementioned sub-sample of 
$\sim$4,000 stars in Fig~\ref{agefe}.\footnote{The `upturn' towards 
high-metallicities at young ages in the GCS sample is likely traced to 
the very young Fm/Fp stars which are difficult to characterise with 
Stromgren photometry alone \citep{Holm09}.}  

It is worth re-emphasising that we are using the \citet{Holm09} variant 
of the GCS solely as a useful `comparator' against which to contrast our 
various MDF metrics / higher-order moments.  It should not be 
interpreted as an endorsement of one solar neighbourhood MDF over 
another; there is a rich literature describing the various pros and cons 
of any number of potential selection biases within this (or any other) 
re-calibration of the GCS \citep[e.g.][]{SB08,Cas11} and we are not 
equipped to enter into that particular debate.  The GCS remains the 
standard-bearer for MDF analysis, reflecting the nature of (fairly) 
volume-limited and (fairly) complete nature, making it ideal for probing 
the active star forming component of the thin disc; other exquisite 
MDFs, including those of the aforementioned (predominantly) thick disc 
\citep{Schlesinger2012} and halo \citep{Sch09} studies, are more suited 
for simulations targeting regions further from the mid-plane than we are 
doing here.  Ideally, of course, we would like to replace the solar 
neighbourhood `comparator' used here (the GCS) with an empirical sample 
more representative of star formation histories associated with massive 
dwarf spiral/irregulars \citep[e.g.][]{Skillman2003,Dolphin03,Kirby11}, but 
until the statistics, completeness, and accuracy of the age and 
metallicity determinations for such distance dwarfs reaches that of the 
solar neighbourhood, we are reluctant to compare (in detail) the 
predictions of the simulations with those of the observations.  Having 
said that, we will comment on, in a qualitative sense, the AMR and MDF 
trends seen in our simulations and how they compare with said dwarfs.

Several key points can be inferred from Fig~\ref{agefe}. First, not 
surprisingly, the metallicities of the stars in the Milky Way's solar 
neighbourhood (GCS) are typically a factor of $\sim$5$-$100$\times$ 
higher at a given age compared with the five simulations.  This reflects 
the discussion of \S\ref{secgal} in relation to the fact that the 
simulations in question are more similar to lower-luminosity disc 
galaxies (in terms of both mass and SFHs), rather than being Milky Way 
`clones'.  The simulations are consistent with the various scaling 
relations to which galaxies adhere \citep{Brook11c}; as such, for their 
mass, their mean metallicities are a factor of $\sim$3$-$5$\times$ lower 
than that of the Milky Way.\footnote{The MDFs and AMRs of systems more 
directly comparable to the Milky Way proper -- i.e., the more massive 
`parent' simulations to those employed here \citep{Stinson10} -- are 
described by \citet{Calura2012} and Bailin et~al. (2012, in prep), 
respectively.  The significant contamination from accreted stars in 
these more massive simulations tends to impact upon both the scatter of 
the AMR and skewness/dispersion of the IMF, in a negative sense, 
relative to the high-feedback models here, for which the accreted 
fraction is negligible.}

More important for our purposes here, there are two additional 
characteristics which are readily apparent in Fig~\ref{agefe}. First, 
the AMR of the solar neighbourhood is essentially non-existent, save for 
a trace of old, metal-poor, stars. In contrast, the corresponding 
regions of the simulations show {\it extremely \rm} correlated AMRs 
(especially those of the fiducial simulations, \tt 11mKroupa \rm and \tt 
11mChab\rm). This is partly traced to the differences in the 
aforementioned SFHs, although the correlation persists (admittedly with 
larger scatter at a given age) even in \tt 11mNoDiff\rm, the simulation 
whose SFH bears the closest resemblance to that of the Milky Way.  The 
impact of these tightly-correlated AMRs manifest themselves 
significantly within the inferred MDFs, a point to which we will return 
in \S\ref{secmdf}.  Qualitatively speaking, these tightly-correlated 
AMRs resemble those predicted by semi-numerical galactic chemical 
evolution models \citep[e.g.][]{Chiap01, Fenner2003, Renda2005, MD05}.

In the bottom right panel of Fig~\ref{agefe}, we also overplot the AMRs 
inferred from the colour-magnitude diagram-derived star formation 
histories of the dwarf irregulars Sextans~A \citep{Dolphin03} and 
IC~1613 \citep{Skillman2003}; like the Milky Way, neither are meant to 
be one-to-one matches to the \tt 11m \rm series of simulations, but in 
some sense they do provide a useful complementary constraint, in the 
sense that their respective star formation histories are not dissimilar 
to those shown in Fig~\ref{sfr} (in particular, those of \tt 
11mKroupa\rm, \tt 11mNoMinShut\rm, and \tt 11mNoRad\rm).  Their 
associated AMRs, while lacking the statistics, completeness, and 
accuracy of the GCS dataset necessary to make detailed quantiative 
comparisons, do show evidence of possessing somewhat stronger 
correlations.  Again, the statistics of these dwarf systems' MDFs and 
AMRs make it difficult to say anything more regarding the degree of 
`agreement' between the \tt 11m \rm series and that encountered in 
nature, but it is suggestive and certainly merits revisting once data 
comparable to that of the GCS becomes available for dwarf 
irregulars/spirals.

Second, the scatter in [Fe/H] at a given stellar age is significantly 
smaller (compared with that of the Milky Way) in the three simulations 
where the injection of thermalised massive star radiation energy to the 
surrounding ISM is included (i.e., \tt 11mKroupa\rm, \tt 11mChab\rm, and 
\tt 11mNoMinShut\rm).  Neglecting this feedback term, within the context 
of these cosmological hydrodynamical disc simulations, acts to increase 
the scatter in [Fe/H], at a given in time, to a level comparable to that 
seen in Milky Way's solar neighbourhood.\footnote{A secondary byproduct 
is also a mildly steeper radial abundance gradient, although the effect 
is minor - recall, Table~\ref{tab}.}  Not surprisingly, the one 
simulation for which metal diffusion was suppressed (\tt 11mNoDiff\rm) 
possesses the largest scatter in [Fe/H] at a given age, particularly at 
early times/low metallicities, where the neglect of diffusion is most 
problematic (again, a point to which we return in \S\ref{secmdf}).

\section{Metallicity Distribution Functions}
\label{secmdf}

Having been informed by the empirical and simulated solar 
neighbourhoods' SFHs and AMRs (\S\ref{secgal} and \S\ref{secagemet}, we 
now present the [Fe/H] metallicity distribution functions (MDFs) for the 
same regions.\footnote{We confirmed that our conclusions are robust to 
the specific definition of the `solar neighbourhood', by increasing its 
vertical range from $\pm$0.5~kpc to $\pm$2~kpc. Similarly, varying the 
radial range from 3.50$\pm$0.25 disc scalelengths, by $\pm$1 scalelength 
has negligble impact (recall from Table~\ref{tab} that the metallicity 
gradients here are shallow).}  Fig~\ref{mdf} shows the MDFs (black 
histograms) for the five simulations, the Milky Way (GCS: lower right 
panel) {and the Local Group dwarf Fornax (also, lower right 
panel, from \citet{Kirby11}). 
The two sub-samples of the GCS are shown; in black, 
the aforementioned (\S\ref{secagemet}) sub-sample of $\sim$4,000 stars 
(matching those shown in Fig~\ref{agefe} -- i.e., the `cleaned' 
sub-sample, but without any distance constraint applied, labeled `GCS' 
in the lower-right panel), and in blue, the volume-limited sample (i.e., 
those lying within 40~pc of the Sun, labeled `GCScut').  As stressed 
earlier, the shape characteristics of the GCS MDF are not contingent 
upon this latter cut; the labels `GCS' and `GCScut' will be employed to 
differentiate between the two, where relevant.  Overlaid in each panel 
is simple `best-fit' (single) Gaussian to the respective distributions 
(and their associated full-width at half-maximum (FWHM) values).
For the Fornax dwarf, we use the full sample of 675 stars taken from 
\citet{Kirby11}, in order to show (perhaps) the best determined
MDF for a representative local dwarf. Three caveats should be noted, 
in relation to the latter: (i) the sample size is 1-2 orders of 
magnitude smaller than the GCS, not surprisingly, considering
the challenging nature of this observational work; (ii) no analogous
`solar neighbourhood' can be identified within this dataset (it is 
simply all the stars in the sample covering a range of fields in
Fornax); and (iii) the uncertainty in [Fe/H] for a given individual
star in Fornax is $\sim$0.5~dex, compared with the $\sim$0.1~dex
associated with individual stars in the GCS.  Fornax is neither
better nor worse than the GCS, as a comparator, so it is useful
to at least show both, as they represent the state-of-the-art, 
observationally-speaking.

Even before undertaking any quantitative analysis of the MDFs, it is 
readily apparent that the simulations (particularly, \tt 11mKroupa\rm, 
and \tt 11mChab\rm) possess an excess of stars to the 
left (i.e., to the negative side) of the peak of the MDF, relative to 
the right, when compared with that of the 
GCS and Fornax (i.e., the simulated MDFs 
are more negatively skewed). This `excess' of lower-metallicity stars 
are formed {\it in situ \rm} during the first $\sim$4~Gyrs of the 
simulations.  The exception to this trend is \tt 11mNoMinShut\rm, for 
which the lack of significant star formation at early epochs (recall, 
Fig~\ref{sfr}) and the extremely flat AMR at late times 
(Fig~\ref{agefe}) conspires to present the narrow and symmetric MDF 
shown in Fig~\ref{mdf}. As noted in \S\ref{secagemet}, for both \tt 
11mNoRad \rm and \tt 11mNoDiff\rm, the larger scatter in [Fe/H] at a 
given age manifests itself in the broader MDFs seen in Fig~\ref{mdf}.

It is worth delving deeper into the source of the broader MDF seen in,
for example, \tt 11mNoRad\rm, relative to the fiducial \tt 11mKroupa\rm.
Here, it is at high-redshift that the radiation energy has an impact 
on the regulation of star formation.  \tt 11mNoRad \rm 
has higher star formation at
early times (Fig~1), but not at later times, primarily because it exhausts 
its available gas, whereas with the radiation energy star formation is
regulated during
that crucial period when gas accretion is at its most active; this gas
remains available at later times to form stars, resulting in the MDF
of \tt 11mNoRad \rm being broader relative to the fiducial. Ultimately,
the length of time that gas spends in the disk before
it forms stars shapes the MDF `width' here. 
With radiation energy included, this gas is in the disk
for a longer period of time, meaning more 
metal mixing occurs.  Linking back to the star formation histories
of Fig~1, we note that most of the gas is
accreted during the first $\sim$6~Gyr, and one can see that
the star formation rate shows that early peak in the case of
\tt 11mNoRad \rm (and \tt 11mNoDiff\rm), but
not in the cases which include radiation energy - i.e., gas that forms
stars (relatively) rapidly after accretion does not mix as much, and
hence the broader MDF.




\begin{figure}
\centerline{ 
\psfig{file=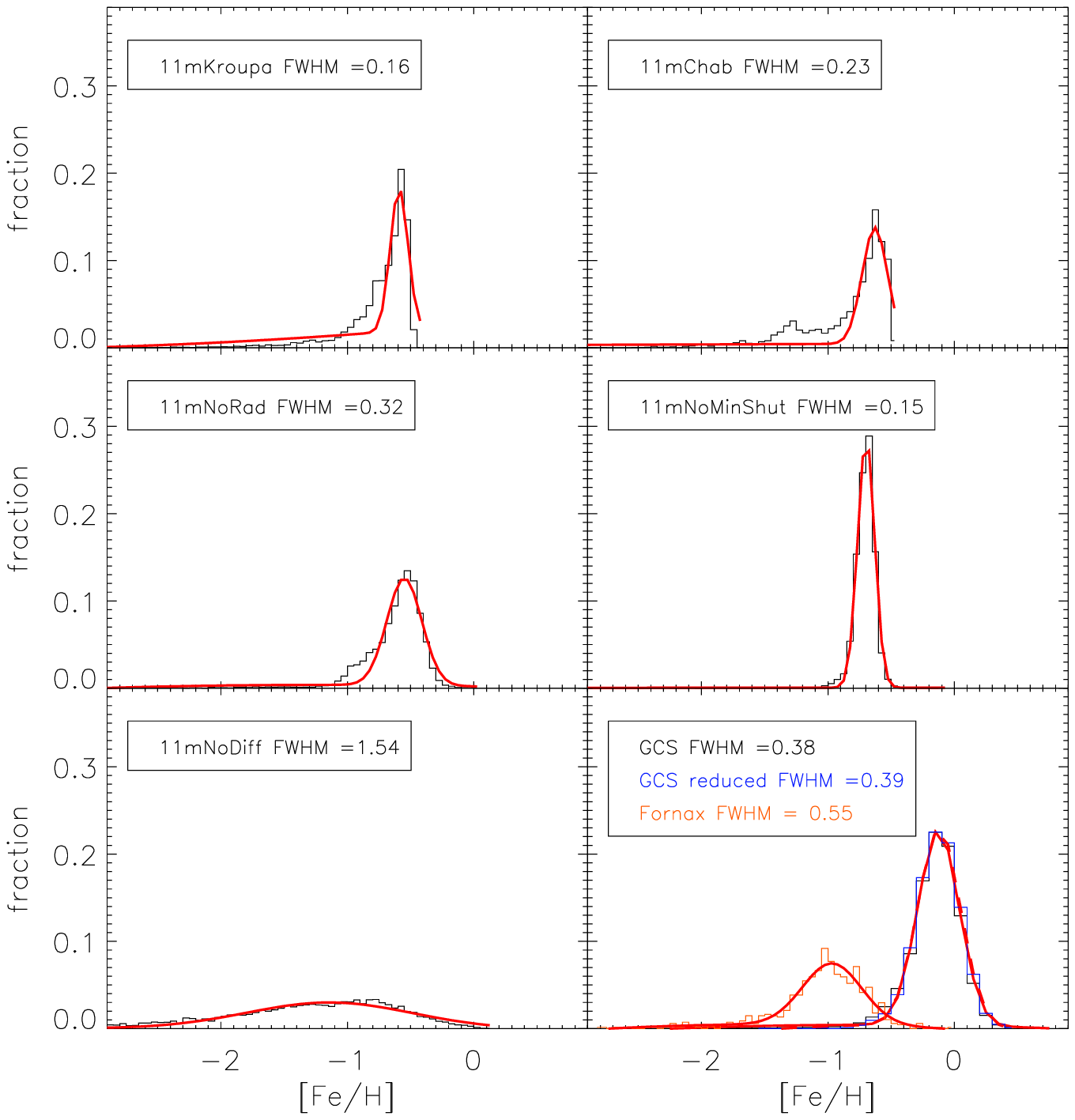,width=9.0cm} 
} 
\caption{The [Fe/H] metallicity distribution functions in the solar 
neighourhoods of the five simulations employed here. The bottom-right
panel shows the MDF of the Milky Way's solar neighbourhood, based upon 
two sub-samples of stars selected from \citet{Holm09},
as well as that for Local Group dwarf Fornax, from \citet{Kirby11} 
(see text for details). In each panel, the overlaid curve is the best-fit
single component Gaussian to the aforementioned MDF; the associated FWHM
of said Gaussian is listed in the inset to each panel.}
\label{mdf}
\end{figure} 

%
\begin{table*}
\centering
\begin{tabular}{|l|l|l|l|l|l|l|}
\hline
Simulation/Dataset & Skewness& Kurtosis& IQR  & IDR  & ICR  & ITPR\\ \hline\hline
\tt 11mKroupa\rm   & $-$1.84($-$1.21) & 3.83(2.59)  & 0.30(0.54) & 0.67(1.13) & 1.59(2.72) & 2.49(4.34)\\ \hline
\tt 11mChab\rm     & $-$1.56($-$1.15) & 2.43(2.37)  & 0.41(0.60) & 0.85(1.28) & 1.71(2.96) & 2.38(5.04)\\ \hline
\tt 11mNoRad\rm    & $-$1.13($-$0.93) & 2.45(1.88)  & 0.26(0.47) & 0.52(0.92) & 1.44(2.07) & 2.39(3.73)\\ \hline
\tt 11mNoMinShut\rm& $+$0.47($-$0.29) & 0.94(0.57)  & 0.13(0.48) & 0.26(0.93) & 0.69(1.79) & 1.97(3.26)\\ \hline
\tt 11mNoDiff\rm   & $-$0.91($-$1.29) & 0.91(2.32)  & 0.96(1.25) & 1.85(2.44) & 3.49(5.18) & 5.06(8.03)\\\hline
\tt GCS\rm         & $-$0.61 & 2.04  & 0.23 & 0.48 & 1.26 & 2.63\\\hline
\tt GCScut\rm      & $-$0.37 & 0.78  & 0.24 & 0.45 & 0.94 & 1.43\\\hline
\tt Fornax\rm      & $\quad\qquad$($-$1.33) & $\qquad$(3.58)  & $\qquad$(0.38) & $\qquad$(2.25) & $\qquad$(2.75) & $\qquad$(2.85)\\\hline
\end{tabular}
\caption{Primary MDF shape characteristics for the solar neighbourhoods 
of the five simulations described here, 
the two sub-samples based upon the 
\citet{Holm09} \tt GCS \rm empirical dataset are as described in the 
text and data for the Fornax dwarf galaxy taken from \citet{Kirby11}. 
After \citet{Fenner2003}, the simulated MDFs were convolved with either
a 0.1~dex Gaussian (left-most entry within each column) or a 0.5~dex
Gaussian (right-most / bracketed entry within each column), 
to mimic the typical uncertainties associated with the 
[Fe/H] determinations in nature (the GCS in the case of the former, and Fornax
in the case of the latter). Column (1): the name of the simulation 
or empirical dataset; Column (2): the skewness of the MDF (5$\sigma$ 
clipping of outliers was imposed, to minimise their impact on the 
determination); Column (3): the kurtosis of the MDF, again with the 
adoption of 5$\sigma$ clipping; Columns (4) $-$ (7): the interquartile 
(IQR), interdecile (IDR), intercentile (ICR), and inter-tenth-percentile 
(ITPR) for each MDF.
}
\label{table}
\end{table*}

We next undertook a quantitative analysis of the MDFs shown in 
Fig~\ref{mdf}, including a determination of the skewness, kurtosis, and 
widths at a range of inter-percentiles of the distributions.  These 
determinations are listed in Table~\ref{table}.  As both skewness and 
kurtosis are highly sensitive to the presence of outliers, we imposed a 
fairly standard 5$\sigma$ clipping to the distributions.  To mimic
the observational uncertainties associated with the determination of
individual stellar [Fe/H] abundances, after \citet{Fenner2003},
the `theoretical' MDFs shown in
Fig~\ref{mdf} were convolved first with either a 0.1~dex Gaussian 
(to mimic the GCS uncertainties - Holmberg et~al. 2009) or a
0.5~dex Gaussian (to mimic the uncertainties with the Fornax data - 
Kirby et~al. 2011).   In Table~\ref{table}, each column has two
numbers; the first is the relevant metric, as measured on the MDF
convolved with a 0.1~dex Gaussian, while the second (in brackets) is
that measured on the MDF convolved with a 0.5~dex Gaussian.
As the simulated MDFs are typically much broader than the GCS 
uncertainties, the impact of the 0.1~dex smoothing is minimal.

As inferred from the above qualitative discussions of the MDF and the 
AMR (\S\ref{secagemet}), MDFs of the simulated solar neighbourhoods are 
all (save for \tt 11mNoMinShut\rm, whose exceedingly flat AMR results in 
the elimination of essentially all tails, positive or negative of the 
MDF's peak) more negatively skewed than that of the Milky Way's solar 
neighbourhood (from both the volume-limited \tt GCScut \rm sample of 
stars, and the unrestricted \tt GCS \rm sample)
and the sample from Fornax. It must be emphasised though that
the typical 0.5~dex uncertainty associated with the determination of
[Fe/H] for individual stars in Fornax means that broadening the 
simulated MDFs, with their typical dispersions of $\sim$0.1~dex, 
by 0.5~dex, `washes out' much of our ability to compare and
contrast the higher-order
MDF metrics, and hence the analysis which follows emphasises the differences
between the simulated MDFs and that of the GCS.  The `tail' of stars to 
the negative side of the peak should not be associated immediately with 
the traditional `G-dwarf problem', since these fully cosmological 
simulations relax the `closed-box' framework which is the hallmark of 
this problem.  Instead, as noted earlier, it is the tightly-correlated 
AMRs which are driving the large negative skewness values; these 
AMRs do not resemble that of the Milky Way's solar neighbourhood. The 
different SFHs are certainly part of the difference, but as noted 
earlier, both the fiducial \tt 11mChab \rm and \tt 11mNoDiff \rm show 
SFHs not dissimilar to the exponentially-declining one of the Milky Way, 
and the coordinated AMRs remain responsible for the larger negative 
skewness in both cases.  An analysis of the kurtosis values for each 
distribution are consistent with this picture. Specifically, the 
simulations' kurtosis values are all larger than those of \tt GCScut\rm, 
and as noted in \S\ref{secgal}, large kurtosis values are driven in part 
by the presence of a `peaky' MDF, but more importantly, the impact of 
extended, `heavy', tails.  These tails (postive or negative) are driven 
by the coordinated AMRs and are reflected in the generally large values 
of kurtosis relative to the Milky Way's distribution.

Alongside the skewness and kurtosis determinations, we present four 
measures of the shape of the MDF, through its dispersion, or width, at 
different amplitudes.  This is done via the width of the inter-quartile 
range (IQR), inter-decile range (IDR), inter-centile range (ICR), and 
the inter-tenth-percentile range (ITPR).\footnote{The IQR corresponds to 
the difference in metallicity between the 25\% lowest metallicity stars 
and the 25\% higher metallicity stars; similarly, the IDR corresponds to 
the difference between the 10\% lowest and 10\% highest metallicity 
stars; etc.}

The metrics associated with these width measures require some comment 
in relation to the information provided by Fig~\ref{mdf}.  Specifically, 
the best-fit single Gaussian fits overlaid in each panel show that 
grossly speaking, the Milky Way's and Fornax's MDF are broader than 
those associated 
with the simulations.\footnote{Save for \tt 11mNoDiff\rm, as noted in 
\S\ref{secagemet}.}  At first glance, the IQR, ITR, etc, measures listed 
in Table~\ref{table} appear counter to this result (which are all, 
essentially, larger than the values found for \tt GCScut\rm, for 
example).  It is important to remember though that, much like the case 
for skewness and kurtosis, these measures of the breadth of the MDF are 
sensitive to the impact of outliers in the tails of the distribution.

It is particularly useful to note the quantitative impact of the role of 
metal diffusion in setting the width of the MDF in tails of the 
distribution.  For example, in the solar neighbourhood of the Milky Way, 
the range in metallicity between the bottom and top 0.1\% of the stars 
is $\sim$2~dex.  For our simulation in which metal diffusion was 
neglected (\tt 11mNoDiff\rm), the corresponding width is $\sim$5~dex -- 
i.e., a factor of $\sim$1000$\times$ greater than the other simulations 
with diffusion and that encountered in the Milky Way, similar to what 
we found for other low diffusion runs \citep{Pilk10}.

After \citet{Cas11}, we show in Fig~\ref{agemdf} the MDF for the solar 
neighbourhood of one of our fiducial simulations (\tt 11mKroupa\rm), but 
now binned more finely in metallicity and colour-coded by age.  Here, 
young stars correspond to those formed in the last 1~Gyr at redshift 
$z$=0; intermediate-age stars are those with ages between 5 and 7~Gyrs; 
old corresponds to stars with ages greater than 9~Gyrs. Using the \tt 
GCS\rm, \citet{Cas11} conclude that the younger stars have a narrower 
MDF that the older stars, consistent with our results (and to be 
expected, given its AMR). \citet{Cas11} also found though that the 
locations of the peaks associated with these old and young stars were at 
the same metallicity, which is not consistent with our simulations.  
Again, this is to be expected given the tightly-correlated AMRs of the 
simulations, relative to that of the Milky Way.

While it may be the case that we are not capturing all of the relevant 
stellar migration physics within these simulations (e.g., bars, spiral 
arms, resonances, etc.), there {\it is \rm} radial migration occurring.  
That said, the radial gradients are shallow for these fiducial dwarfs 
($-$0.01$-$0.02~dex/kpc, recalling Table~\ref{tab}\footnote{Flatter than 
the gradients seen in our work on the massive galactic analogues to 
these dwarfs \citep{Pilkington2012}, consistent with the empirical 
work on gradients in dwarfs \citep[e.g.][]{Carr08}.}) and, as such, over 
the few kpcs of `disc' associated with each simulated dwarf, systematic 
migration of metal-rich inner-disc stars outwards (and vice versa) has 
little impact on the position of the MDF `sub-structure' (in which the 
young, intermediate, and old `peaks' are offset by $\sim$0.3$-$0.5~dex 
from one another).  Again, this is entirely consistent with the expected 
behaviour, based upon the AMR Fig~2.

\begin{figure}
\centerline{ 
\psfig{file=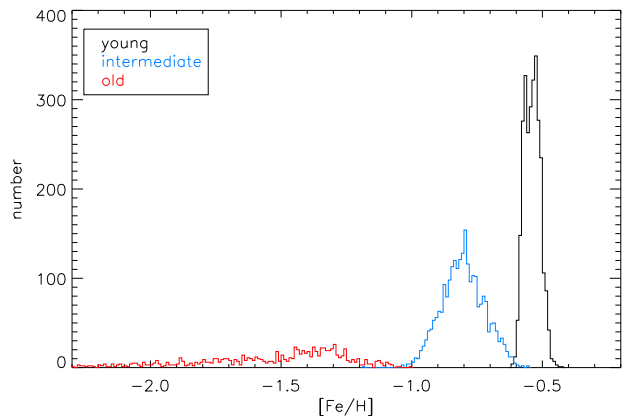,width=9.0cm} 
} 
\caption{The [Fe/H] MDF in the `solar neighbourhood' of \tt 11mKroupa\rm, 
split into three age intervals: young (black) defined as any star 
particle in the solar neighbourhood at redshift $z$=0 with an age less 
than 1~Gyr; intermediate (blue) defined as any star with an age between 
5 and 7~Gyr; old (red) defined as any star with an age greater than 
9~Gyr.}
\label{agemdf}
\end{figure} 

\begin{figure}
\centerline{ 
\psfig{file=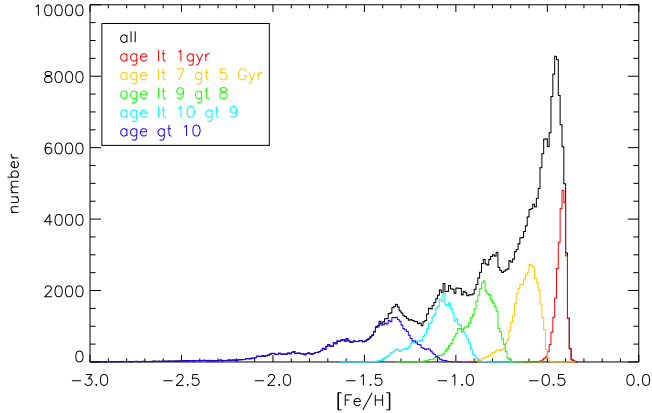,width=9.0cm}
} 
\caption{The [Fe/H] MDF of the `bulge' of \tt 11mKroupa\rm; here, the 
bulge is simply defined as those stars located within 2~kpc of the 
galactic centre at $z$=0. Alongside the full MDF (black line), 
sub-components based upon the age intervales noted in the inset are 
overdrawn.}
\label{agemdf2}
\end{figure} 

The central regions of our simulations show similar characteristics to 
those seen in the simulated solar neighbourhoods.  Specifically, the 
[Fe/H] MDF of the `bulge' (inner 2~kpc) shows a peak near 
[Fe/H]$\sim$$-$0.5, with a number of sub-components at lower metallicity 
which correspond to progressively older and metal-poor populations (see 
Fig~\ref{agemdf2}).  In spirit, such behaviour has been seen in the MDF 
of the bulge of the Milky Way, where \citet{Ben11} finds two 
populations, also separated comparably in age and metallicity, to which 
they associate seaprate formation scenarios. Similarly, \citet{Hill11} 
finds bulge sub-components within the MDF which they also separate into 
separate age, metallicity, and kinematic sub-structures, concluding the 
metal-poor component can be associated with an old spheroid, and the 
more metal-rich component can be associated with a longer timescale 
event (perhaps the evolution of the bar / psudeo-bulge).  In our 
simulations, we see systematic trends in age and kinematics for each 
metallicity sub-component of Fig~5, in the sense of the more metal-poor 
components being older and progressively less rotationally-supported, in 
exactly the manner one might predict from the AMR (\S\ref{secagemet}).  
It should be emphasised though that within the simulations, the 
behaviour of these age, metallicity, and kinematic `sub-structure' in 
the bulge MDF is continuous, rather than showing any discrete transition 
from rotational support to anisotropic velocity support.

Finally, we now examine in slightly more detail the behaviour of the 
extreme metal-poor tails of the simulated MDFs (see Figs~\ref{cummdfall} 
and \ref{cummdf}).  In Fig~\ref{cummdfall}, we show all stars beyond the 
inner 3~kpc (and within 10~kpc), in order to minimise the effect of the 
`spheroid' stars in the analysis.  We experimented, as before, with the 
impact of using a full kinematic decomposition between disc and spheroid 
stars, but again, for these dwarfs, the spatial cut alone is 
indistinguishable from the decomposed galaxy.  In Fig~\ref{cummdf}, we 
only show those star particles lying within the previously defined 
`solar neighbourhoods' of each simulation.

One additional curve is included in both figures (labeled \tt 109CH\rm), 
that of the disc generated with the adaptive mesh refinement code 
\textsc{Ramses} and described by \citet{San09}, in which diffusion is 
handled `naturally'.  As noted previously, each of the \tt 11m \rm 
series of simulations employ the \citet{SWS10} metal diffusion framework 
with a diffusion coefficient C=0.05, except for (obviously) \tt 
11mNoDiff\rm which assumes C=0.0.

Each of the cumulative MDFs (Figs~\ref{cummdfall} and \ref{cummdf}) are 
normalised. In both cases, the normalisation occurs at the [Fe/H] 
corresponding to the metallicity of the lowest 1\% of the stars (in 
terms of [Fe/H]).  For plotting purposes, these are then aligned 
arbitrarily at [Fe/H]$\equiv$+0.0, to show the relative distributions of 
extremely low-metallicity stars within each simulation and the empirical 
datasets.  One could take a different approach and, say, normalise at 
(i) the same metallicity, (ii) the same amplitude, or (iii) the same 
number of stars.  For example, in our analysis of the \citet{Gov10} 
bulgeless dwarf galaxy simulations \citep{Pilk10}, we adopted (i), 
normalising all MDFs at [Fe/H]=$-$2.3. This was similar in spirit to 
\citet{Sch09}, who fixed the normalisations of the Milky Way halo and 
Local Group dwarf spheroidal MDFs to be unity at the metallicity 
corresponding to the lowest (in terms of [Fe/H]) $\sim$100 stars in 
each.  For distributions which peak at (potentially) very different 
metallicities, such normalisations can result in significant outliers 
which are not necessarily driven by any MDF `tail'.\footnote{In the case 
of the analysis of \citet{Sch09}, the similarity of the positions of the 
peaks of the Milky Way halo and Local Group dSph MDFs meant that their 
analysis was robust against the choice of normalisation.}  For our work 
here, while small quantitative differences exist depending upon the 
adopted normalisation, the qualitative results are robust regardless of 
the choice.

\begin{figure}
\centerline{ 
\psfig{file=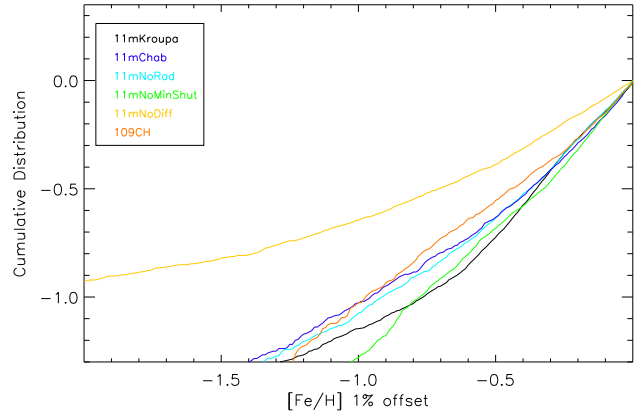,width=9.0cm} 
} 
\caption{The cumulative MDFs of the five \tt 11m \rm simulations: \tt 
11mKroupa \rm (black), \tt 11mNoRad \rm (cyan), \tt 11mChab \rm (blue), 
\tt 11mNoMinShut \rm (green), and \tt 11mNoDiff \rm (gold), in addition 
to that of \tt 109CH \rm (orange: \citet{San09}). For these 
six simulations, all stars lying within 3 and 10~kpc of their respective 
galactic centres are included in the analysis. The normalisation in each 
case is at the metallicity corresponding to that of the lowest 1\% (in 
terms of [Fe/H]) of the stars in each case.}
\label{cummdfall}
\end{figure} 

What is immediately clear from even a cursory examination of 
Fig~\ref{cummdfall} is that the relative distribution of extremely 
metal-poor stars within all the simulations in which metal diffusion 
acts - i.e., all but \tt 11mNoDiff \rm - are consistent with each other. 
This reflects graphically what we have commented upon earlier in 
relation to the tabulated ICR and ITPR values for the various MDFs 
(Table~\ref{table}). Specifically, the lack of metal diffusion 
within \tt 11mNoDiff \rm drives its discrepant ICR and ITPR values 
(Table~\ref{table}), and its outlier status in Fig~\ref{cummdfall}.  
When compared with Fig~5 of \citet{Pilk10}, one can see that the overly 
`heavy' metal-poor tail to the MDF of \tt 11mNoDiff \rm matches that 
encountered in, for example, the low metal-diffusion simulations of 
\citet{Gov10}.\footnote{Demonstrating the quantitative power of the MDF 
to constrain the magnitude of diffusion within SPH simulations of galaxy 
formation.}  One fairly robust conclusion that can be drawn from
Fig~\ref{cummdfall} is that the \it relative \rm distribution of 
\it extremely \rm metal-poor stars is robust against the choice of feedback
scheme; instead, diffusion plays a more important role in shaping
this distribution.

\begin{figure}
\centerline{ 
\psfig{file=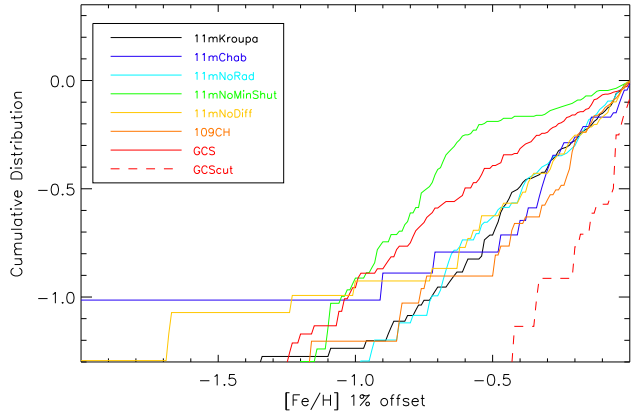,width=9.0cm} 
} 
\caption{The cumulative MDFs of the analogous solar neighbourhoods 
associated the five \tt 11m \rm simulations: \tt 11mKroupa \rm (black), 
\tt 11mNoRad \rm (purple), \tt 11mChab \rm (blue), \tt 11mNoMinShut \rm 
(green), and \tt 11mNoDiff \rm (gold), in addition to that of \tt 109CH 
\rm (orange: \citet{San09}). For these six simulations, 
the solar neighbourhood is defined spatially to include stars lying 
between 3 and 3.5 disc scalelengths from their respective galactic 
centres, and within 0.5~kpc of the mid-plane. The \tt GCS \rm and \tt 
GCScut \rm sub-samples described in \S\ref{secgal} are shown in red.  
The normalisation for each curve is as described for 
Fig~\ref{cummdfall}.}
\label{cummdf}
\end{figure} 

In some sense, the better `statistics' afforded by Fig~\ref{cummdfall} 
provides a `cleaner' picture than that seen when restricting the 
analysis to just the `solar neighbourhoods'.\footnote{And given the lack 
of any substantial gradient in the stellar populations for these dwarfs, 
the comparison is not invalid.}  For completeness though, in 
Fig~\ref{cummdf} we also show the cumulative MDFs of the metal-poor 
tails for each dataset, normalised as in Fig~\ref{cummdfall}.  We should 
emphasise though that the small number of star particles in the `bottom' 
1\% (in terms of metallicity) of the \tt 11mNoDiff\rm, \tt 11mChab\rm, 
and \tt GCScut \rm samples ($\sim$30 in each) make any interpretation 
susceptible to small-number statistics (and stochastic point-to-point
`fluctuations' which are `averaged' over when employed the full disc, 
as in Fig~\ref{cummdfall}).

\section{Summary}
\label{summary}

Employing a suite of five simulations of an M33-scale late-type disc 
galaxy, each with the same assembly history, but with different 
prescriptions for stellar and supernovae feedback, initial mass 
functions, metal diffusion, and supernova remnant cooling `shut-off' 
period, we have analysed the resulting chemistry of the stellar 
populations, with a particular focus on the metallicity distribution 
functions and the characteristics of the extreme metal-poor tail of said 
distributions.

In the context of the distribution of metals (in the sense of the 
higher-order moments of the resultings MDFs) within these discs, the
impact of feedback and the IMF is more subtle than that of, for 
example, metal diffusion.  Employing a \citet{Chabrier2003} IMF, rather
than the \citet{Kroupa1993} form adopted in our earlier work, does
impact significantly on the resulting star formation history (and 
associated, reduced, stellar mass fraction, resulting in 
remarakably close adherence to a wade range of empirical 
scaling relations - Brook et~al. (2011c)).

The star formation histories of the `solar' neighbourhoods associated 
with each simulation show exceedingly tight age-metallicity relations. 
In shape, these relations are akin to those predicted by classical 
galactic chemical evolution models \citep[e.g.][]{Chiap01, Fenner2003}, 
but bear somewhat less resemblance to that seen, for example, in the 
Milky Way's solar neighbourhood \citep{Holm09}.  These correlated 
age-metallicity relations result inexorably in (negatively) skewed MDFs 
with large kurtosis values, when compared with the Milky Way. Star 
formation histories of dwarf irregulars, which qualitatively speaking 
are a better match to those of the \tt 11m \rm series of simulations 
presented here, suggest though that somewhat steeper age-metallicity 
relations might eventuate in nature in these environments 
\citep[e.g.][]{Dolphin03,Skillman2003}.  MDFs and AMRs of a comparable 
quality to that of the GCS \citep{Holm09} will be required to 
subtantively progress the field.

An excess `tail' of extremely metal-poor stars (amongst the bottom 
0.1$-$1\% of the most metal-poor stars) -- $\sim$2$-$3~dex below the 
peak of the MDF -- exists in all of the simulations, as reflected in 
their inter-centile (ICR) and inter-tenth-of-a-percentile (ITPR) region 
measures.  This tail is particularly problematic in simulations without 
metal diffusion (\tt 11mNoDiff\rm) and those for which the diffusion 
coefficient was set relatively low \citep[e.g.][]{Gov10, Pilk10}.  As 
demonstrated, the ICR and ITPR, in the absence of metal diffusion, can 
be $\sim$30$-$3000$\times$ larger than that encountered in the Milky 
Way.

We end with a re-statement of our initial caveat.  The simulations 
presented here (particularly the fiducials, \tt 11mKroupa \rm and \tt 
11mChab\rm) have been shown to be remarkably consistent with a wide 
range of scaling relations \citep{Brook11c}.  That said, their star 
formation histories are more akin to those of NGC~6822, Sextans A, WLM, 
and to some extent, the LMC (at least in the case of \tt 11mChab\rm) -- 
i.e., these systems are not `clones' of the Milky Way. We have used the 
wonderful Geneva-Copenehagen Survey's wealth of data to generate 
empirical age-metallicity relations and metallicity distribution 
functions against which to compare, but exact one-to-one matches are not 
to be expected.  That said, they do provide useful, hopefully generic, 
relations against which to compare.  In the future, we hope to extend 
our analysis to equally comprehensive datasets for the LMC, making use 
of, for example, the data provided by the Vista Magellanic Cloud Survey 
\citep{Cioni2011}.

\section*{Acknowledgments} 
KP acknowledges the support of STFC through its PhD Studentship 
programme (ST/F007701/1). BKG, CBB, and RJT acknowledge the support of 
the UK's Science \& Technology Facilities Council (ST/F002432/1 \& 
ST/H00260X/1). BKG and KP acknowledge the generous visitor support 
provided by Saint Mary's University and Monash University. We thank the 
DEISA consortium, co-funded through EU FP6 project RI-031513 and the FP7 
project RI-222919, for support within the DEISA Extreme Computing 
Initiative, the UK's National Cosmology Supercomputer (COSMOS), and the 
University of Central Lancashire's High Performance Computing Facility.
The helpful guidance of the anonymous referee is gratefully 
acknowledged.

\bibliographystyle{mn2e}
\bibliography{11mJeans}

\label{lastpage}

\end{document}